\documentclass[twoside,twocolumn,9pt]{article}
\usepackage{extsizes}
\usepackage[super,sort&compress,comma]{natbib} 
\usepackage[version=3]{mhchem}
\usepackage[left=1.5cm, right=1.5cm, top=1.785cm, bottom=2.0cm]{geometry}
\usepackage{balance}
\usepackage{mathptmx}
\usepackage{sectsty}
\usepackage{graphicx} 
\usepackage{lastpage}
\usepackage[format=plain,justification=justified,singlelinecheck=false,font={stretch=1.125,small,sf},labelfont=bf,labelsep=space]{caption}
\usepackage{float}
\usepackage{fancyhdr}
\usepackage{fnpos}
\usepackage[english]{babel}
\addto{\captionsenglish}{%
  
}
\usepackage{array}
\usepackage{droidsans}
\usepackage{charter}
\usepackage[T1]{fontenc}
\usepackage[usenames,dvipsnames]{xcolor}
\usepackage{setspace}
\usepackage[compact]{titlesec}
\usepackage{hyperref}
\usepackage{cleveref}
\usepackage{bm}

\usepackage{epstopdf}

\definecolor{cream}{RGB}{222,217,201}

\begin{document}

\pagestyle{fancy}
\thispagestyle{plain}
\fancypagestyle{plain}{
\renewcommand{\headrulewidth}{0pt}
}

\makeFNbottom
\makeatletter
\renewcommand\LARGE{\@setfontsize\LARGE{15pt}{17}}
\renewcommand\Large{\@setfontsize\Large{12pt}{14}}
\renewcommand\large{\@setfontsize\large{10pt}{12}}
\renewcommand\footnotesize{\@setfontsize\footnotesize{7pt}{10}}
\makeatother

\renewcommand{\thefootnote}{\fnsymbol{footnote}}
\renewcommand\footnoterule{\vspace*{1pt}%
\color{cream}\hrule width 3.5in height 0.4pt \color{black}\vspace*{5pt}} 
\setcounter{secnumdepth}{5}

\makeatletter 
\renewcommand\@biblabel[1]{#1}            
\renewcommand\@makefntext[1]%
{\noindent\makebox[0pt][r]{\@thefnmark\,}#1}
\makeatother 
\renewcommand{\figurename}{\small{Fig.}~}
\sectionfont{\sffamily\Large}
\subsectionfont{\normalsize}
\subsubsectionfont{\bf}
\setstretch{1.125} 
\setlength{\skip\footins}{0.8cm}
\setlength{\footnotesep}{0.25cm}
\setlength{\jot}{10pt}
\titlespacing*{\section}{0pt}{4pt}{4pt}
\titlespacing*{\subsection}{0pt}{15pt}{1pt}

\fancyfoot{}
\fancyfoot[LO,RE]{\vspace{-7.1pt}\includegraphics[height=9pt]{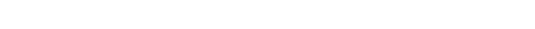}}
\fancyfoot[CO]{\vspace{-7.1pt}\hspace{13.2cm}\includegraphics{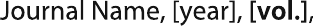}}
\fancyfoot[CE]{\vspace{-7.2pt}\hspace{-14.2cm}\includegraphics{head_foot/RF}}
\fancyfoot[RO]{\footnotesize{\sffamily{1--\pageref{LastPage} ~\textbar  \hspace{2pt}\thepage}}}
\fancyfoot[LE]{\footnotesize{\sffamily{\thepage~\textbar\hspace{3.45cm} 1--\pageref{LastPage}}}}
\fancyhead{}
\renewcommand{\headrulewidth}{0pt} 
\renewcommand{\footrulewidth}{0pt}
\setlength{\arrayrulewidth}{1pt}
\setlength{\columnsep}{6.5mm}
\setlength\bibsep{1pt}

\makeatletter 
\newlength{\figrulesep} 
\setlength{\figrulesep}{0.5\textfloatsep} 

\newcommand{\topfigrule}{\vspace*{-1pt}%
\noindent{\color{cream}\rule[-\figrulesep]{\columnwidth}{1.5pt}} }

\newcommand{\botfigrule}{\vspace*{-2pt}%
\noindent{\color{cream}\rule[\figrulesep]{\columnwidth}{1.5pt}} }

\newcommand{\dblfigrule}{\vspace*{-1pt}%
\noindent{\color{cream}\rule[-\figrulesep]{\textwidth}{1.5pt}} }

\makeatother

\twocolumn[
  \begin{@twocolumnfalse}
{\includegraphics[height=30pt]{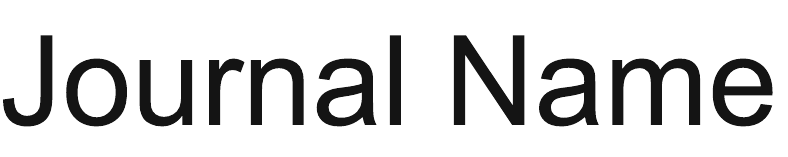}\hfill\raisebox{0pt}[0pt][0pt]{\includegraphics[height=55pt]{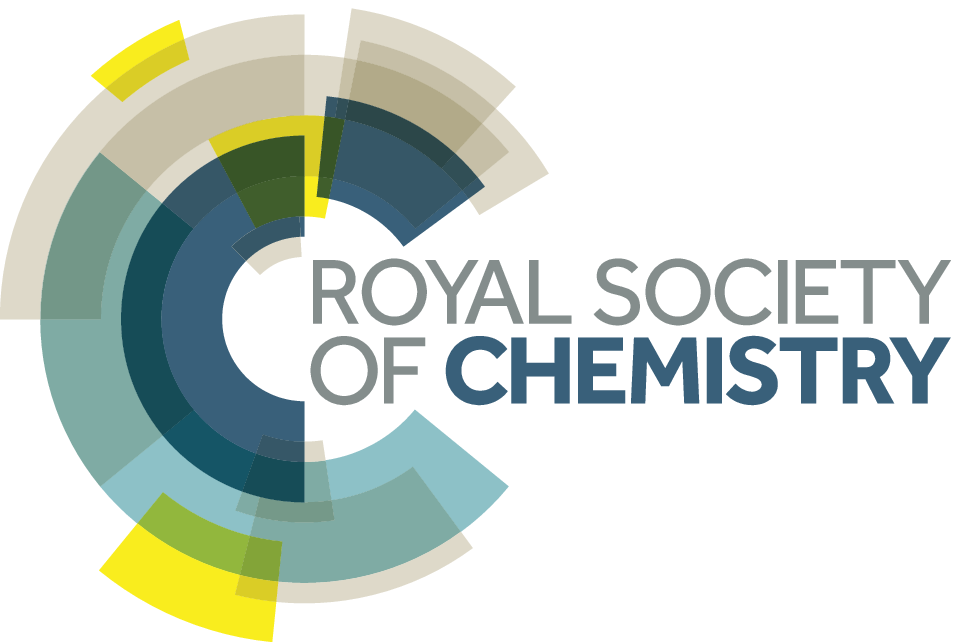}}\\[1ex]
\includegraphics[width=18.5cm]{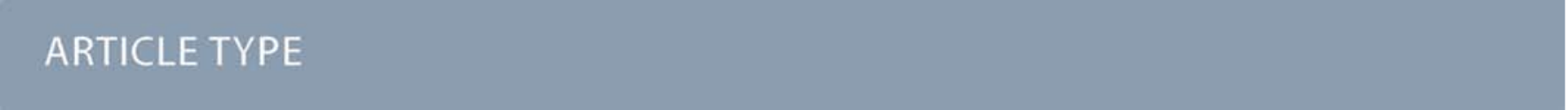}}\par
\vspace{1em}
\sffamily
\begin{tabular}{m{4.5cm} p{13.5cm} }

\includegraphics{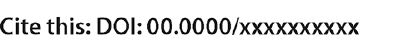} & \noindent\LARGE{\textbf{A statistical and machine learning approach to the study of astrochemistry}} \\
\vspace{0.3cm} & \vspace{0.3cm} \\

 & \noindent\large{Johannes Heyl,\textit{$^{a\ddag}$} Serena Viti,\textit{$^{b, a\ddag}$} and Gijs Vermariën\textit{$^{b}$}} \\

\includegraphics{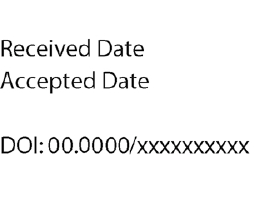} & \noindent\normalsize{In order to obtain a good understanding of astrochemistry, it is crucial to better understand the key parameters that govern grain-surface chemistry. For many chemical networks, these crucial parameters are the binding energies of the species. However, there exists much disagreement regarding these values in the literature. In this work, a Bayesian inference approach is taken to estimate these values. It is found that this is difficult to do in the absence of enough data. The Massive Optimised Parameter Estimation and Data (MOPED) compression algorithm is then used to help determine which species should be prioritised for future detections in order to better constrain the values of binding energies. Finally, an interpretable machine learning approach is taken in order to better understand the non-linear relationship between binding energies and the final abundances of specific species of interest.} \\

\end{tabular}

 \end{@twocolumnfalse} \vspace{0.6cm}

  ]

\renewcommand*\rmdefault{bch}\normalfont\upshape
\rmfamily
\section*{}
\vspace{-1cm}


\footnotetext{\textit{$^{a}$~Department of Physics and Astronomy, University College London, Gower Street, WC1E 6BT, London, UK; E-mail: johannes.heyl.19@ucl.ac.uk}}
\footnotetext{\textit{$^{b}$~Leiden Observatory, Leiden University, PO Box 9513, 2300 RA Leiden, The Netherlands }}
\footnotetext{\ddag These authors contributed equally.}



\section{Introduction}
Giant Molecular Clouds in our Milky Way as well as in other galaxies host gas which is almost entirely molecular, with densities above $\sim$ 100 cm$^{-3}$ and temperature below $\sim$ 100 K.
These denser, cooler regions contain a significant fraction of the non-stellar baryonic matter in a galaxy and they are usually much more massive than large tenuous ones.
The importance of these regions lie in the fact that they are key for our understanding of how galaxies form and evolve because this denser, cooler gas is the reservoir of matter that forms stars and planets, as well as the gas that fuels the centres of galaxies.

From an astrochemical point of view, due to their high densities and low temperatures, these regions are great laboratories to study the interactions of gas and dust, with
species from the gas phase ‘freezing’ onto the dust grains present, and forming icy mantles rich in hydrogenated as well as complex organic molecules (COMs), due to the many fast surface reactions that take place. As stars form in these clouds (or if any other energetic process takes place) then the dust temperature may reach
the mantle sublimation temperature ($\sim$ 100K), and the molecules in the mantles are injected into the gas, where they react and form new, more complex, molecules. Associated with star formation, as well as with active galactic nucleus (AGN) activity, are highly supersonic collimated jets and molecular outflows. When the outflowing material encounters the quiescent gas of molecular cloud, it creates shocks, where the grain mantles are (partially) sputtered and the refractory grains are shattered. Again, here, the interaction of gas and dust varies within very short timescales and the effects of chemistry and dynamics are interlocked in a complex non-linear fashion. In summary,
 the gas and dust surface compositions exhibit a complicated time dependent, non-linear chemistry that strongly depends on the physical environment. There are many open questions - still - about such interaction: what is the unprocessed ice composition? What are the efficiencies of the viable surface reactions? And how do the energetics of the ISM (cosmic rays, UV radiation, shocks) influence the processed ices? In order to determine accurate estimates of the abundances of molecular species  as a function of all the parameters that influence their chemistry we need to be able to answer such questions. In other words, we need to understand the chemical pathways towards each molecule and its dependencies on the density, temperature and energetics of the gas and dust before molecules can be truly considered powerful tools. 
 
In recent years coupling chemical and radiative transfer models for the interpretation of molecular emission has been routinely done  and the success of such techniques  has varied to different degrees, depending on whether one wants to model the physical and chemical structure, or the hydrodynamical history of the gas\cite{Bisbas_2014, Viti_2014, Kazandjian_2016,Huang_2022}. However the shortcomings of such methods are two-fold: (i) understanding the physical conditions in molecular gas via a systematic and applicable to many galaxies methodology is an inverse problem subject to complicated chemistry that varies non-linearly with both time and the physical environment\cite{Makrymallis_2014}; hence it may not have a solution, solutions might
not be unique and/or might not depend continuously on the
observational data. Traditionally astrochemistry has always been dominated by trial and error grid-based analysis combined with simple statistics\cite{Lefevre_2014}, an approach that becomes impossible or ineffective when datasets (e.g from ALMA) and/or parameter space are large, complex, or heterogeneous; (ii) the knowledge of the micro-physics and chemistry of what occurs on the dust is well behind what is known for the gas-phase. While surface reactions and dynamics (including desorption and diffusion)  can be
experimentally investigated (but always within a constrained range of laboratory conditions), experimental data for interstellar ices are still
limited. In order to make the best use of
experimental resources, the chemical data that models require need
to be prioritized according to what will have the most impact.
 
In recent years progress based on the use of Bayesian as well as Machine Learning (ML) techniques to deal with both the issues above has been made, from the creation of neural network based statistical  emulators\cite{Damien_paper, chemulator, Grassi_autoencoders} in order to optimize the integration of chemical, radiative transfer and hydrodynamical models to the use of ML techniques to disentangle multiple gas components in unresolved beams\cite{Damien_NMF}. 

In this paper we will focus our attention to Bayesian and ML techniques applied to the study of chemical networks and the key parameters that govern their interactions. In recent years there has been a substantial body of work concentrating on reducing the cost of solving chemical networks computations using various techniques from Monte Carlo approaches to constrain important reactions\cite{holdship}, to automated reduction schemes\cite{Grassi_2012, Xu_2019}, to topological methods\cite{Grassi_2013, Heyl1} to, finally, ML algorithms\cite{Grassi_autoencoders, TangandTurk}. In parallel several studies have concentrated on the estimation of poorly known reaction rates, with particular emphasis on surface chemical networks: an initial approach considered a simple grain-surface network and applied a Bayesian inference method coupled with Markov Chain Monte Carlo sampling in order to infer reaction rates\cite{holdship}. This was followed up with an approach that considered the topological structure of the network\cite{Heyl1}, while another exploited the characteristics of the chemical reaction mechanism to significantly reduce the dimensionality of the problem under consideration by simply considering the binding energies and the role they play in the determination of grain-surface chemistry\cite{Heyl2}. Subsequent work using the 'Massive Optimised Parameter Estimation and Data compression' (MOPED) algorithm, helped make predictions about which ice species needed to be detected to reduce the variance of binding energy estimates \cite{Heyl_MOPED}.

Due to the significant role that binding energies play in grain-surface chemistry, we shall concentrate on the estimation of binding energies as well as on priorization of the ice species that should be observed with instruments such as the JWST to better improve our understanding of their values. We will then use machine learning interpretability to consider the forward relationship between binding energies and the abundances of species of interest. Our methods are described in Sections 2. The results are presented in Section 3  and a brief conclusion is given in Section 4.

\section{Methodology}
In this section we first describe the chemical code we use and the chemical assumptions we make  in our work (Section 2.1), followed by a description of the analytical approach we employ (Section 2.2). 

\subsection{The Chemical Code}

All modelling in this work is done with the open-source astrochemical code UCLCHEM \cite{UCLCHEM}\footnote{https://uclchem.github.io/}. The chemistry of a collapsing dark cloud was modelled. The dark cloud collapsed isothermally at 10K from $10^{2}$ cm$^{-3}$ to $10^{6}$ cm$^{-3}$ over 5 million years. The composition of the ices as a result of the ensuing chemistry was then compared to the recent ice observations with the James Webb Space Telescope (JWST)  \cite{McClure}.

As this work focuses solely on grain-surface chemistry, it is pertinent to describe the details of the underlying reaction mechanisms we consider in this work. This will be used as justification to explain why binding energies are of such great importance in the context of this work. 

In UCLCHEM, the main grain-surface reaction mechanism is the Langmuir–Hinshelwood mechanism \cite{Hasegawa}. The rate at which two species A and B react through diffusion is given by: 

\begin{equation}\label{reaction_rate}
k_{AB} = \kappa_{AB}\frac{(k^{A}_{hop}+k^{B}_{hop})}{N_{site}n_{dust}},
\end{equation}

where $N_{site}$ is the number of sites on the grain surface and $n_{dust}$ is the dust grain number density.

In equation \ref{reaction_rate}, $k^{X}_{hop}$ is the thermal hopping rate of species $X$ on the grain surface which is given as: 

\begin{equation}\label{hopping_rate_equation}
k^{X}_{hop} = \nu_{0}\exp\left(-\frac{E_{D}}{T_{gr}}\right),
\end{equation}

where $E_{D}$ is the diffusion energy of the species, $T_{gr}$ is the grain temperature and $\nu_{0}$ is the characteristic vibration frequency of species $X$. The diffusion energy is a fraction of the binding energy of the species, $E_{b}$. 

The characteristic vibration frequency, $\nu_{0}$, is defined as: 

\begin{equation}
\nu_{0} = \sqrt{\frac{2k_{b}n_{s}E_{b}}{\pi^{2}m}},
\end{equation}

where $k_{b}$ is the Boltzmann constant, $n_{s}$ is the grain site density and $m$ is the mass of species. 




The final term, $\kappa_{AB}$, which gives the reaction probability is:

\begin{equation}\label{kappa}
\kappa_{AB} = \max\left(\exp{\left(-\frac{2a}{\hbar}\sqrt{2\mu k_{b}E_{A}}\right)}, \exp{\left(-\frac{E_{A}}{T_{gr}}\right)}\right),
\end{equation}

where $\hbar$ is the reduced Planck constant, $\mu$ is the reduced mass, $E_{A}$ is the reaction activation energy, $k_{b}$ is Boltzmann's constant and $a = 1.4$ Angstrom is the thickness of a quantum mechanical barrier that is used as the default in UCLCHEM. The reaction probability encodes the competition between the quantum mechanical probability of a tunnelling through a rectangular barrier of thickness $a$, which is the first term, and the thermal reaction probability, which is the second term.  

Species do not necessarily need to react with each other on the grains. It is also possible for them to diffuse away from a potential reactant or evaporate. As such,  a modification needs to be made to the $\kappa_{AB}$ term to take this into account. This is the reaction-diffusion competition \citep{Chang, GarrodandPauly}. The reaction probability is now defined as: 

\begin{equation}
\kappa_{AB}^{final} = \frac{p_{reac}}{p_{reac} + p_{diff} + p_{evap}},
\end{equation}

where $p_{reac}$, $p_{diff}$ and $p_{evap}$ represent the probabilities of species A and B reacting, diffusing and evaporating per unit time, respectively. These quantities are defined as: 

\begin{equation}
p_{reac} = \max(\nu_{0}^{A}, \nu_{0}^{B})\kappa_{AB},
\end{equation}
\begin{equation}
p_{diff} = k_{hop}^{A} + k_{hop}^{B}
\end{equation}

and

\begin{equation}
p_{evap} = \nu_{0}^{A}\exp\left(-\frac{E_{b}^{A}}{T_{gr}}\right) + \nu_{0}^{B}\exp\left(-\frac{E_{b}^{B}}{T_{gr}}\right).
\end{equation}

The term $\kappa_{AB}$ in Equation \ref{reaction_rate} is replaced with $\kappa_{AB}^{final}$ .

Equations 1-8 show that the key quantities are $\nu_{0}$, $k_{hop}^{X}$, $E_{b}$ and $E_{A}$. The first two are functions of the binding energies of the reacting species. We assume that the activation energies in Equation \ref{kappa} are well-known, so do not include these as parameters to be estimated. 

If we wish to better understand grain-surface diffusion-based chemistry, we must have accurate values of the binding energies of species. For most cases, at 10K, the reactant with the lower binding energy will dominate the total hopping rate, due to the exponential dependence of the hopping rate on the diffusion energy. Across the literature, there is often significant disagreement when it comes to the values of binding energies\cite{UMIST, Wakelam, Quenard}. While there exist many different methods of estimating these values\cite{experimental_approach, Ferrero, Villadsen}, we utilise a Bayesian inference approach.

The chemical network used consists of a gas-phase and ice-phase network. The gas-phase network is the UMIST network\cite{UMIST}. The ice network used is the same as in previous work\cite{Heyl_MOPED}, but augmented with a sulphur network based on work done to explain the sulphur depletion problem\cite{Laas_Caselli}. The inclusion of the sulphur network is important, since recently sulphur-bearing species have been confirmed in the ices\cite{McClure}.
\subsection{Analytical Approach}
\subsubsection{Bayesian Inference}

\begin{table}
\setlength\tabcolsep{2.0pt}
\hspace*{1.5cm}
 \begin{tabular}{||c c||} 
 \hline
 Species & Abundances relative to H \\ [1ex] 
 \hline\hline
 H$_{2}$O & $(8.8 \pm 1.1) \times 10^{-5}$  \\ 
 \hline
 CO & $(2.2 \pm 0.3) \times 10^{-5}$ \\
 \hline
 CO$_{2}$ & $(1.1 \pm 0.2) \times 10^{-5}$\\
 \hline
 CH$_{3}$OH & $(3.1 \pm 0.7) \times 10^{-6}$\\
 \hline
  NH$_{3}$ & $(8.8 \pm 1.6) \times 10^{-6}$\\
 \hline
 CH$_{4}$ & $(1.8 \pm 0.1) \times 10^{-6}$\\
 \hline
  OCN & $\sim 2.0 \times 10^{-7}$\\
 \hline
  SO$_2$ & $\sim 6.6 \times 10^{-8}$\\
 \hline
   OCS & $\sim 1.3 \times 10^{-7}$\\
 \hline

\end{tabular}
\caption{The abundances and uncertainties taken from \citet{McClure}. These abundances were taken from sources with an $A_v$ of 95.}
\label{abundance_table}
\end{table}

One of the goals of this work is to estimate the binding energies of the most diffusive species in the network. These species were chosen based on a literature search that suggested they were amongst the species with the lowest values for their binding energies. The binding energy parameters are represented as a vector, $\boldsymbol{E}$ = (E$_{b, \mathrm{H}}$, E$_{b,\mathrm{H}_{2}}$, E$_{b,\mathrm{C}}$, E$_{b,\mathrm{CH}}$, E$_{b,\mathrm{N}}$, E$_{b,\mathrm{CH}_{3}}$, E$_{b, \mathrm{NH}}$, E$_{b,\mathrm{CH}_{4}}$, E$_{b,\mathrm{O}}$). UCLCHEM was rewritten so that it would take the vector as an input and output the abundances of species of interest. The mapping between the input and output can be summarised as $\boldsymbol{Y} = f(\boldsymbol{E})$, where $f$ represents UCLCHEM. We are looking to estimate the binding energies that give us abundances that match our measurements best. This is an inverse problem, as we are trying to determine the best-fitting inputs that give an output of interest. 

Bayes' Law was used to solve this inference problem. Given the data, $\boldsymbol{d}$, of abundance measurements of species, the probability distributions of the binding energies of interest are given by: 

\begin{equation}
P(\boldsymbol{E} \vert \boldsymbol{d}) = \frac{P(\boldsymbol{d} \vert \boldsymbol{E})P(\boldsymbol{E})}{P(\boldsymbol{d})},
\end{equation}
where $P(\boldsymbol{E} \vert \boldsymbol{d})$ is the posterior probability distribution, $P(\boldsymbol{E})$ is the prior, $P(\boldsymbol{d} \vert \boldsymbol{E})$ is the likelihood and $P(\boldsymbol{d})$ is referred to as the evidence. The prior distribution encodes the initial understanding of the binding energy distribution. The likelihood gives the data's likelihood as a function of the binding energies. Within the likelihood function, the physical model is encoded. The evidence serves as a normalising factor and represents the marginalised likelihood. The posterior distribution represents the updated probability distribution of reaction rates based on the data, the prior distribution, and the physical model.

The prior for all binding energies was selected as a uniform distribution between 400 K and 2000 K. The abundance measurements, given in Table \ref{abundance_table}, were assumed to be Gaussian. The species without associated uncertainty, $\mathrm{OCN}$, $\mathrm{SO}_2$ and $\mathrm{OCS}$, were given a relative uncertainty of 50\%. Assuming a Gaussian distribution, the likelihood function can be specified:

\begin{equation}\label{likelihood}
\centering
P(\boldsymbol{d} \vert \boldsymbol{E}) =  \prod_{i=1}^{n_{d}} \frac{1}{\sqrt{2\pi}\sigma_{i}} \exp\left({-\frac{(d_{i}-Y_{i})^{2}}{2\sigma_{i}^{2}}}\right),
\end{equation}

where $n_{d}$ is the number of observations and $\sigma_{i}$ is the uncertainty of the $i$th observation. Only the species for which there are abundances measurements are indexed over. 

The inference was implemented using the UltraNest Python package \citep{UltraNest}. The package implements efficient methods to construct a neighbourhood to sample from, allowing for better convergence of the sampling of the likelihood \citep{Buchner1, Buchner2}. The package conveniently also outputs the maximum likelihood-estimator, $\boldsymbol{E_{ML}}$, which will be utilised later.

\subsubsection{The MOPED Algorithm}
While our knowledge of the molecular inventory in the gas-phase is quite complete, we are still far from being confident about the ice composition as well as the ice chemistry. To this end, we employ the "Massive Optimised Parameter Estimation and Data compression" (MOPED) algorithm \citep{Heavens, Heavens2017, Heavens2020ExtremePhysics}. 

The aim of the algorithm is to determine which of the $M$ species in our chemical network would best constrain our knowledge for our $p$ binding energy parameters. In this work, $p = 9$ and $M = 119$. Some binding energies will have a greater influence on certain species than others. The key is to determine the species that are most sensitive to the binding energies of interest. In doing so, we can then make recommendations for future ice observations as was done in a proof-of-concept work recently \cite{Heyl_MOPED}.  

There will be uncertainty associated with all of our potential future abundance measurements. It is likely that these uncertainties will vary by species. However, it is difficult to determine these species-dependent uncertainties a priori. As such, we assume the uncertainty on each abundance measurement is the same. This is summarised in a covariance matrix: $\boldsymbol{C} = diag(\sigma_{1}^{2}, \sigma_{2}^{2}, ... \sigma_{M}^{2})$. 

We apply a filtering technique developed by \citet{Heavens, Heavens2017, Heavens2020ExtremePhysics} who propose using a linear combination of the final abundances of network, $\boldsymbol{Y}$, to compress data points into numbers. Such a compression takes the form: 

\begin{equation}
    c_\alpha = \boldsymbol{b_{\alpha}^{T}}\boldsymbol{Y},
    \label{compressed_data_point}
\end{equation}

where $\alpha$ ranges from 1 to $p$ and ${\boldsymbol{b_{\alpha}}}$ is a set of orthonormal linear filters. Each filter vector is unique to each parameter and does not contain information contained in any of the other vectors. $\boldsymbol{Y}$ represents a vector containing the final, steady-state abundances for some value of $\boldsymbol{E}$, though we employ $\boldsymbol{E} = \boldsymbol{E_{ML}}$, which can be obtained from the Bayesian inference, as this has been found to be sufficient as the fiducial model \cite{Heavens,Heavens2017}. For each $c_{\alpha}$, there will be greater dependence on some of the components of ${\boldsymbol{b_{\alpha}}}$ than others. As each component represents a different species, this implies that a component with a greater magnitude has more information about that parameter.

The vectors ${\boldsymbol{b_{\alpha}}}$ are given by

\begin{equation}
    \boldsymbol{b}_1 = \frac{\boldsymbol{C}^{-1} \boldsymbol{Y},_1}{\sqrt{\boldsymbol{Y},_1^T \boldsymbol{C}^{-1} \boldsymbol{Y},_1 }}
    \label{filter_1}
\end{equation}

and 

\begin{equation}
    \boldsymbol{b}_\alpha = \frac{\boldsymbol{C}^{-1} \boldsymbol{Y},_\alpha - \sum_{\beta = 1 }^{\alpha -1}{(\boldsymbol{Y,}_\alpha^T \boldsymbol{b,}_\beta)\boldsymbol{b,}_\beta }}{\sqrt{ \boldsymbol{Y,}_\alpha^T \boldsymbol{C}^{-1} \boldsymbol{Y},_\alpha - \sum _{\beta = 1}^{\alpha -1 }(\boldsymbol{Y},_\alpha^T \boldsymbol{b},_\beta)^2}}, 
    \label{filter_2}
\end{equation}

\begin{figure*}
\includegraphics[width=2.1\columnwidth]{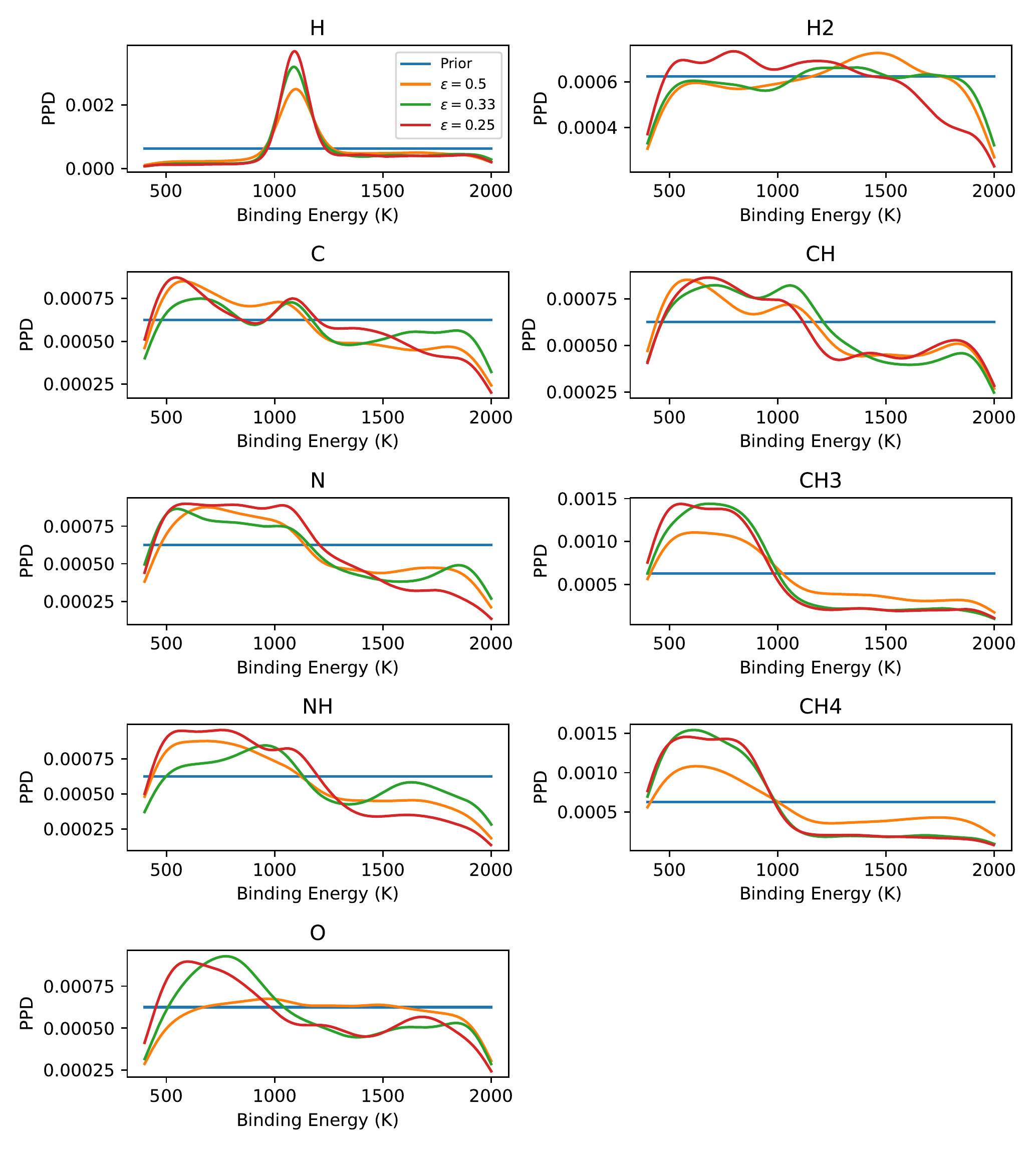}
\caption{Marginalised posterior distributions of the binding energies of the diffusive species we consider of interest in this work. We also plot the uniform prior distribution. Only H's binding energy marginalised posterior distribution differs significantly from the prior distribution. For the other binding energies, there is less difference. This is due to the lack of enough sufficiently constraining data. We also observe that decreasing the value of $\epsilon$ in general decreases the variance of the distribution. Both of these points motivate the need for further ice observations to reduce the variance of the distributions.  }
\label{bayesian_inference_plots}
\end{figure*}

\begin{figure*}[h!]
\hspace*{-1cm}
\includegraphics[width=2.3\columnwidth]{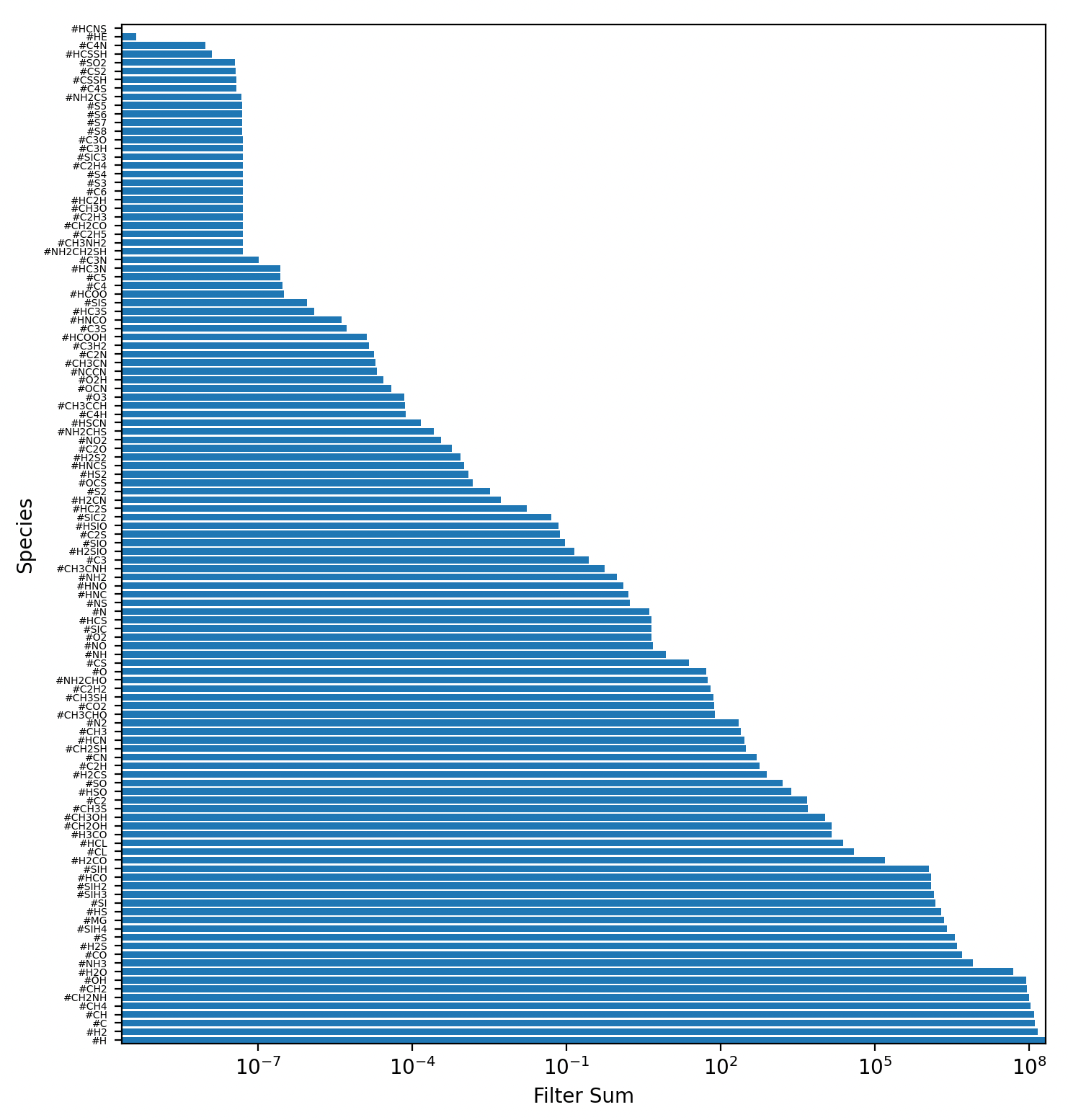}
\caption{Bar chart displaying the filter sums for all grain-surface species. Species with a larger filter sum are higher priority detection targets, as they are more affected by the binding energies of the species we consider. Some of the highest-ranked species have already been detected, which potentially implies that future observations should aim to improve the level of precision of these abundance measurements.}
\label{bar_chart}
\end{figure*}

where $\boldsymbol{Y},_{\alpha}$ is the partial derivative of $\boldsymbol{Y}$ with respect to the parameter $\alpha$ around the point $\boldsymbol{Y}=f(\boldsymbol{E_{ML}})$. The equations for $\boldsymbol{b}_{\alpha}$ were derived via a Lagrange multiplier procedure\cite{Heavens}. When it is said that all filters are orthonormal, this means that 

\begin{equation}
    \boldsymbol{b_{\alpha}^{T}}\boldsymbol{C}\boldsymbol{b_{\beta}} = \delta_{\alpha\beta},
    \label{orthonormality}
\end{equation}

which is another way of saying that all filter vectors are uncorrelated. Each component of $\boldsymbol{b_{\alpha}}$ is weighted:
\begin{itemize}
    \item inversely by the size of the uncertainties associated with each species, as encoded by the covariance matrix
    \item the sensitivity of the species' abundance to the value of the binding energy, which is represented by the $\boldsymbol{Y},_\alpha$.
\end{itemize}


If one wished to obtain a ranking of species in terms of their importance in helping constrain binding energies, one would need to come up with a 'score' for each species.  Recall, that as the magnitude of each component of $\boldsymbol{b_{\alpha}}$ is a weight for that species' influence on the parameter $\alpha$, one would need to sum over the absolute values of the components of $\boldsymbol{b_{\alpha}}$ for species across all $\alpha$. That is, we perform the sum over our linear filters

\begin{equation}
\sum_{\alpha=1}^{p} [|b_{\alpha}^{1}|, |b_{\alpha}^{2}|..., |b_{\alpha}^{M}|].
\end{equation}

We now have a ``filter sum" for each of the $M$ species in our network, which serves as a means of comparing the importance of each species in helping us better constrain binding energy distributions. A species with a larger filter sum will have a larger influence in helping constrain the $p$ binding energies.

\begin{figure*}[h!]
\hspace*{-2cm}
\includegraphics[width=2.5\columnwidth]{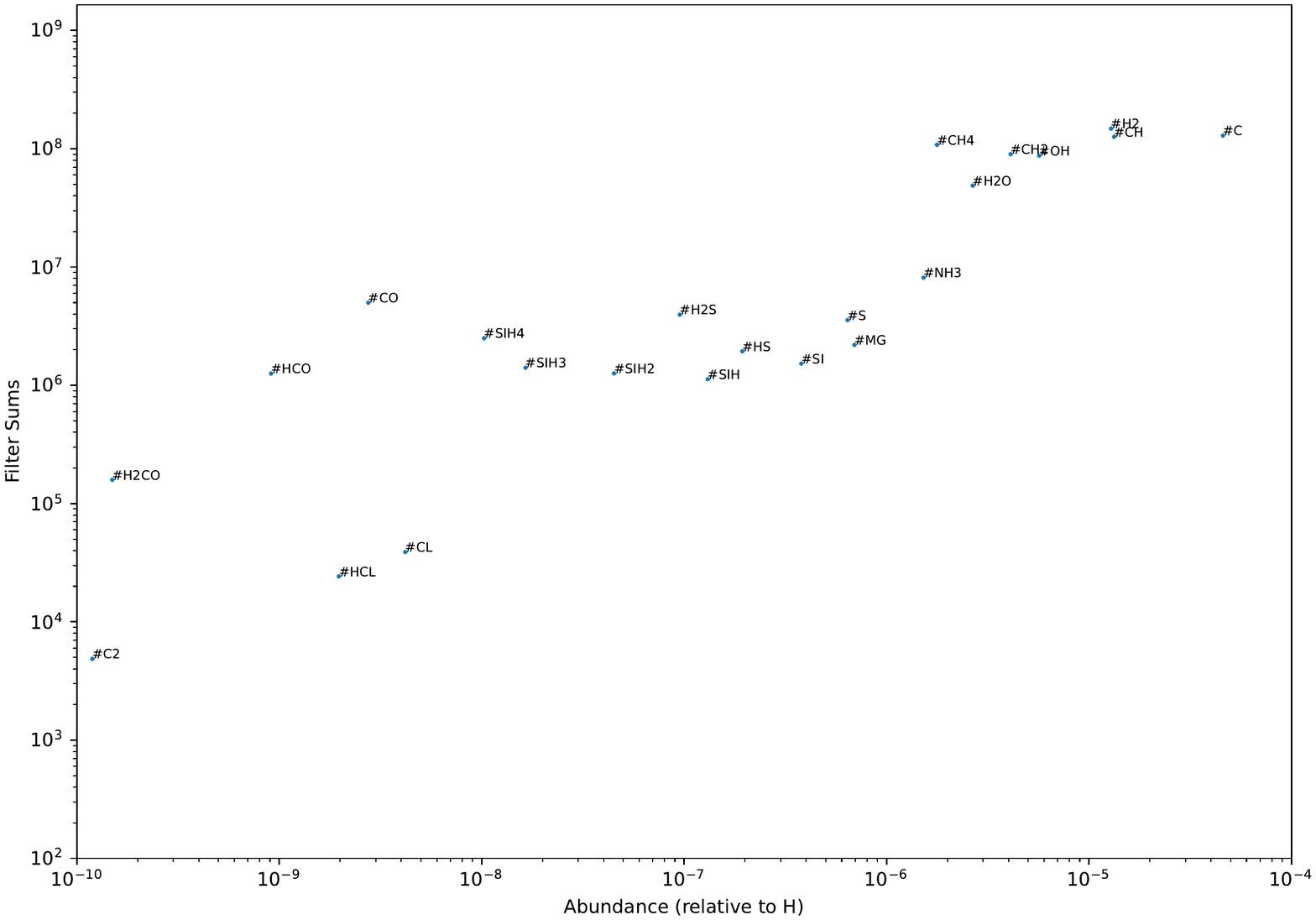}
\caption{Scatter plot depicting filter sum against the predicted abundances when the maximum-likelihod estimate for the binding energies is input into UCLCHEM. Given constraints on instrumental uncertainties, we should look to prioritise species that are not only important, as determined by their filter sums, but that can also be realistically detected. These include saturated species such as \ce{\#CH4}, \ce{\#NH3}, \ce{\#SiH4}, \ce{\#H2S} and \ce{\#H2O}, as well as their precursors. We find that many of the species we observe are the intermediate species formed during the creation of the saturated species in Table \ref{abundance_table}. This indicates that understanding these intermediate products is essential to better constraining the binding energies of interest. }
\label{scatter_plot}
\end{figure*}

\subsubsection{Machine Learning Interpretability}
The previous methods explore the influence of the abundances on the values of the binding energies. This is an inverse problem. In order to tackle the forward problem of assessing the impact of the binding energies on the abundances instead, one needs to use a different set of methods. 

As UCLCHEM solves a system of coupled ordinary differential equations, it stands to reason that the relationship between the input parameters (the binding energies) and the output parameters (the abundances of species) is non-linear. As such, the relationship between the input and output is not necessarily intuitive and is likely to be different for various 'binding energy regimes'. We make use of machine learning interpretability to help uncover this relationship. 

In order to better understand the relationships between the inputs and outputs, we utilise SHapley Additive exPlanations (SHAP) \citep{SHAP_paper}. SHAP approximates Shapley values: these are measures of the marginal contribution of a feature to the output value, relative to the mean value of all output in the dataset \citep{Shapley}. This is done by considering various coalitions of feature values. A coalition of features represents all subsets of the total set of features. The Shapley value of a feature represents the average change in the prediction when that feature is included in the coalition of features selected. This change is assessed by considering the change in the prediction when the feature is included, averaged over all coalitions \citep{molnar2022}. However, this becomes computationally unfeasible as the number of features grows, as the number of subsets grows exponentially with the number of feature. SHAP is particularly useful, as it approximates the Shapley values, greatly reducing the time taken to compute them. This is done through the use of the TreeSHAP algorithm \citep{TreeSHAP}.

500,000 data points were created from UCLCHEM by using a Latin Hypercube sampling scheme \cite{LHS} implemented with the help of the Python surrogate modelling toolbox\cite{LHS_python}. We employ the XGBoost Python package\footnote{https://xgboost.readthedocs.io/en/stable/index.html} to build an XGBoost regressor\cite{XGBoost_original_paper} that is made to fit the relationship between the input parameters and the output abundance for each species. 

\section{Results}
\subsection{Results of the Bayesian Inference}

At first, the Bayesian inference was run using the original dataset. However, it was found that despite running the inference in parallel using MPI over 128 cores, that there was no convergence, even after several days. This was attributed to the fact that the model struggled to match the constraints. Many of these constraints have very low relative error, compared to the data used in previous works which typically had relative errors of the order of 50\% \cite{holdship, Heyl1, Heyl2}. A nested sampler will move from areas of low likelihood to areas of high likelihood. However, if the model struggles to find combinations of parameters that lead to a higher likelihood, then it will inevitably take longer to  perform the inference. To properly run the inference, a significantly larger computing cluster would be required. 
As an alternative, we decided to investigate how the relative error, $\epsilon$, impacted the obtained posterior probability distributions. We used values of 0.5, 0.33 and 0.25 and ran the inference each time. Our results are displayed in Figure \ref{bayesian_inference_plots}. Also plotted are the prior distributions. 

We observe that with the exception of hydrogen's  binding energy,  the binding energy posteriors are prior-dominated. However, it can also be seen that a decrease in the relative error of the data appears to be accompanied by a decrease in the variance of some of the posteriors, such as for \ce{CH3}, \ce{CH4}, \ce{NH} and O. This is consistent with lower variance posteriors for \ce{H} and \ce{O} binding energy with the artificially reduced uncertainties for \ce{H2O} observation in prior work\cite{Heyl_MOPED}.
However, even in this scenario we are finding that our posteriors have a relatively large variance. The best way to address this is to figure out which other species we should observe to further constrain the distributions. 

\subsection{Using the MOPED Algorithm}
We now look to analyse the results of the MOPED algorithm. The fiducial model we use is the one with $\epsilon = 0.25$. In Figure \ref{bar_chart} we plot the filter sums for each species to provide us with an initial ranking. We only consider species formed on the grains. As the UCLCHEM code models both the bulk and the surface abundances,  we sum the abundances of each species on the surface as well as in the bulk to provide us with a total abundance on the ices.

However, in order to inform future ice observations, it would be useful to also consider the likely abundances of each species. Ideally, we would wish to observe species that are highly abundant and that have large filter sums. The first requirement means it is easier to observe a species given a particular instrumental uncertainty, whilst the second ensures that we are observing species that are dependent on the binding energies and are therefore relevant to the chemistry we are considering. To do this, we plot the filter sum of each species against the abundance produced when we use binding energies equal to $\boldsymbol{E_{ML}}$. The resulting plot is Figure \ref{scatter_plot}. We only consider species with an abundance greater than $10^{-10}$ relative to H, as anything less abundant  is unlikely to be detected in the ices. As in previous work\cite{Heyl_MOPED}, we observe that the species \ce{H2O}, \ce{CH4}, \ce{NH3}, \ce{H2S}, \ce{SiH4}, CO and \ce{H2CO} are amongst the highest-ranked species with abundances that are predicted to be detectable. These species all have modes in the range considered by JWST. Unlike in previous work, however, \ce{CO2}, \ce{CH3OH} and \ce{HCN} are not amongst the most significant species. This can be attributed to the fiducial model, as we used different constraints, which lead to the maximum-likelihood estimate being different.

\subsection{Insights from the Machine Learning Interpretability}
Previously, we considered the impact of the data, i.e. the species abundances, on the binding energy values and their distributions. We now wish to consider the opposite situation, which is the impact of the binding energy values on the final steady-state abundances of molecules of interest. This is important to consider as the binding energy of a species can be dependent on the ice-composition as well as on the individual sites\cite{Grassi_binding_energy_distributions, Das_binding_energies}.

\begin{figure*}
\includegraphics[width=2.1\columnwidth]{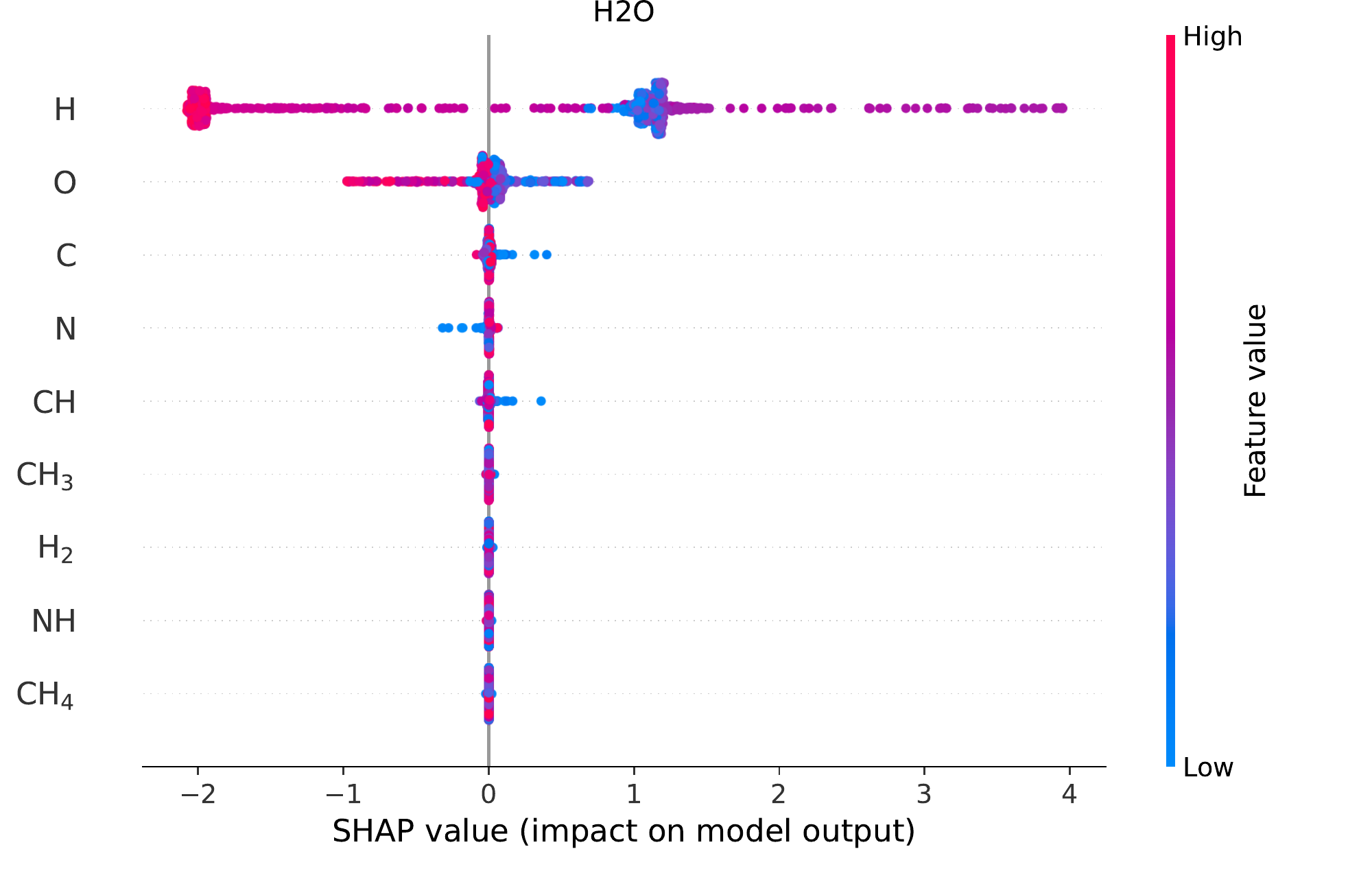}
\caption{A beeswarm plot for the statistical emulator trained to predict \ce{H2O}'s abundance. The features are listed from top to bottom in decreasing order of importance to the model output. Along the horizontal axis, individual predictions are plotted in terms of their SHAP value, that is the change to the log-abundance relative to the average value in the dataset. Recall that the SHAP value states the difference in the value of the model output for that prediction relative to the global average. Furthermore, the points are colour-coded in terms of the size of the feature value. We observe that the binding energies of H and O are the most important features. This makes sense, as both species are necessary to form water via successive hydrogenations of an oxygen atom.}
\label{H2O}
\end{figure*}

\begin{figure*}
\includegraphics[width=2.1\columnwidth]{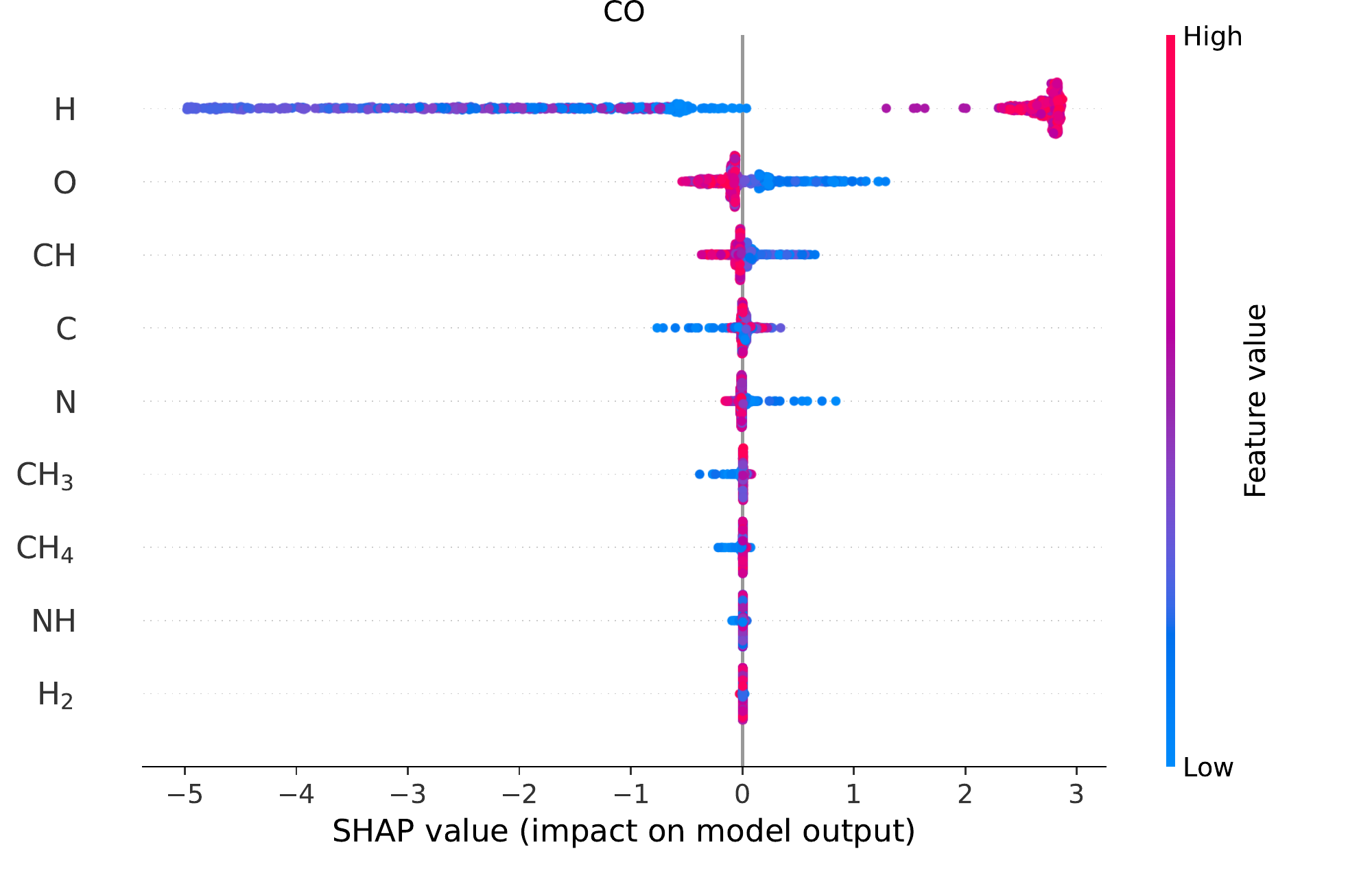}
\caption{A beeswarm plot for the statistical emulator trained to predict \ce{CO}'s abundance. The features are listed from top to bottom in decreasing order of importance to the model output. Along the horizontal axis, individual predictions are plotted in terms of their SHAP value, that is the change to the log-abundance relative to the average value in the dataset. Recall that the SHAP value states the difference in the value of the model output for that prediction relative to the global average. Furthermore, the points are colour-coded in terms of the size of the feature value. We observe that the binding energies of H and O are the most important features. Increasing H's binding energy appears to increase CO's abundance, which can be attributed to a decrease in the efficiency of the hydrogenation of CO.}
\label{CO}
\end{figure*}

In the interest of brevity, we consider a subset of the molecules so as to demonstrate the effectiveness of this approach as a proof-of-concept. We are interested in better understanding the importance of each of the features in predicting the final abundance of a species of interest, as well as the relative importances of the features. Figures \ref{H2O} and \ref{CO} are so-called beeswarm plots for \ce{H2O} and CO respectively. The features are listed from top to bottom in decreasing order of importance to the model output. Along the horizontal axis, individual predictions are plotted in terms of their SHAP value. Recall that the SHAP value states the difference in the value of the model output for that prediction relative to the global average. Furthermore, the points are colour-coded in terms of the size of the feature value. From this, we can attempt to better understand the directionality of each feature's relationship with the output. 

From the beeswarm plots, we can make a number of comments about which binding energies are most relevant for that species. For example, \ce{H2O} is unsurprisingly dependent on the H and O. Others seems less intuitive, such as CO's strong dependence on the H binding energy or \ce{CO2}'s dependence on nitrogen. These can typically be reasoned out by considering the chemical network used. 

\begin{figure*}
\includegraphics[width=2.1\columnwidth]{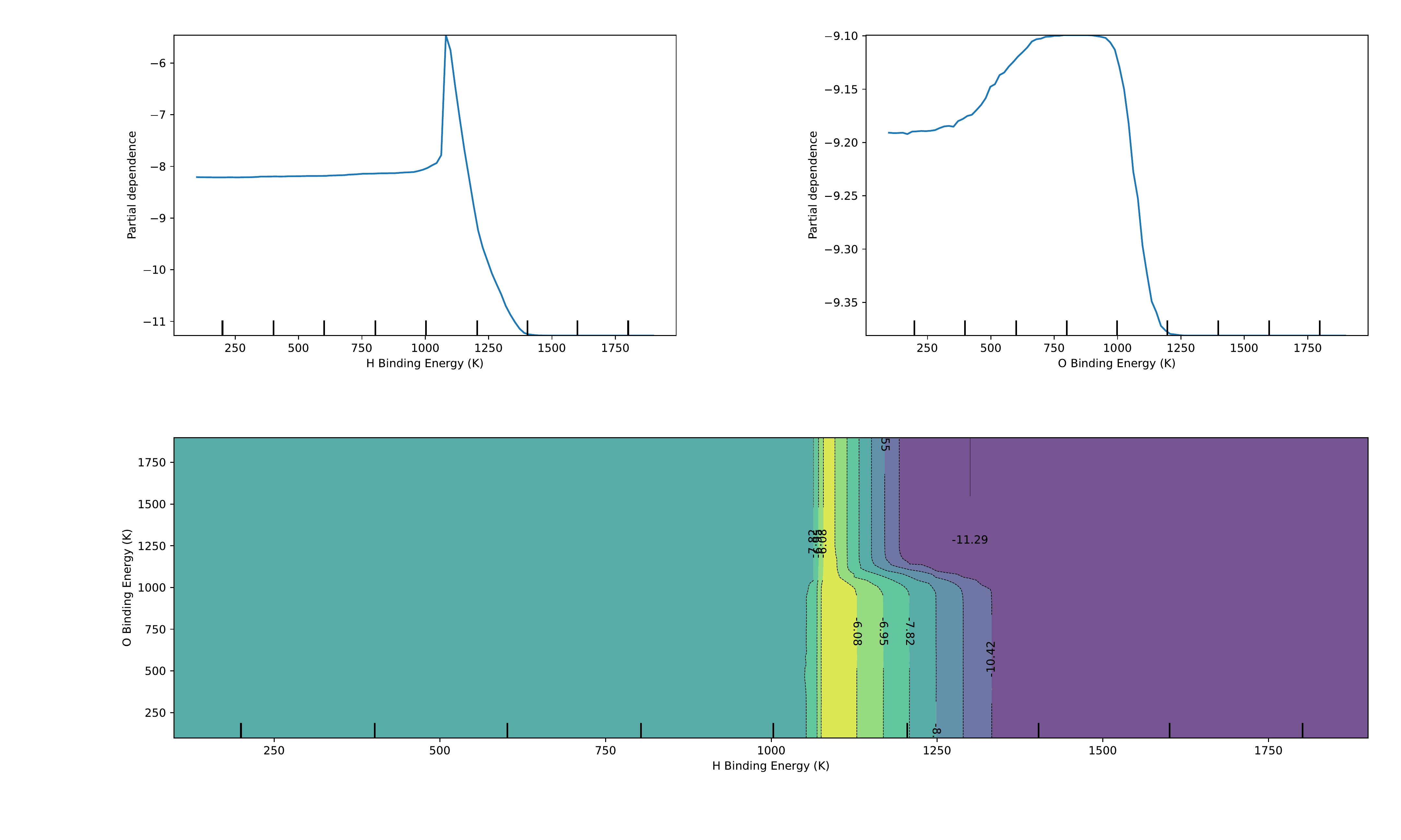}
\caption{Top: A plot of the 1-D partial dependence plots of the binding energies of H and O for water. The partial dependence represents the expected value of the log-abundance of water as a function of the variable in features, marginalised over all other features. We observe that for a narrow range of atomic hydrogen's binding energies at around 1100 K, there is a sharp increase in the abundance of water. This is roughly the point at which the marginalised posterior distribution for H's binding energy in Figure \ref{bayesian_inference_plots} peaks. The dependence for $\ce{O}$'s binding energy shows a similar consistency with the posteriors, having a clear preference for energies smaller than $\sim 1000$K.
Bottom: A 2-D partial dependence plot for the binding energies of H and O. Yellow represents the region with the highest abundance of water.}
\label{H2O partial dependence plots}
\end{figure*}

\begin{figure*}
\includegraphics[width=2.1\columnwidth]{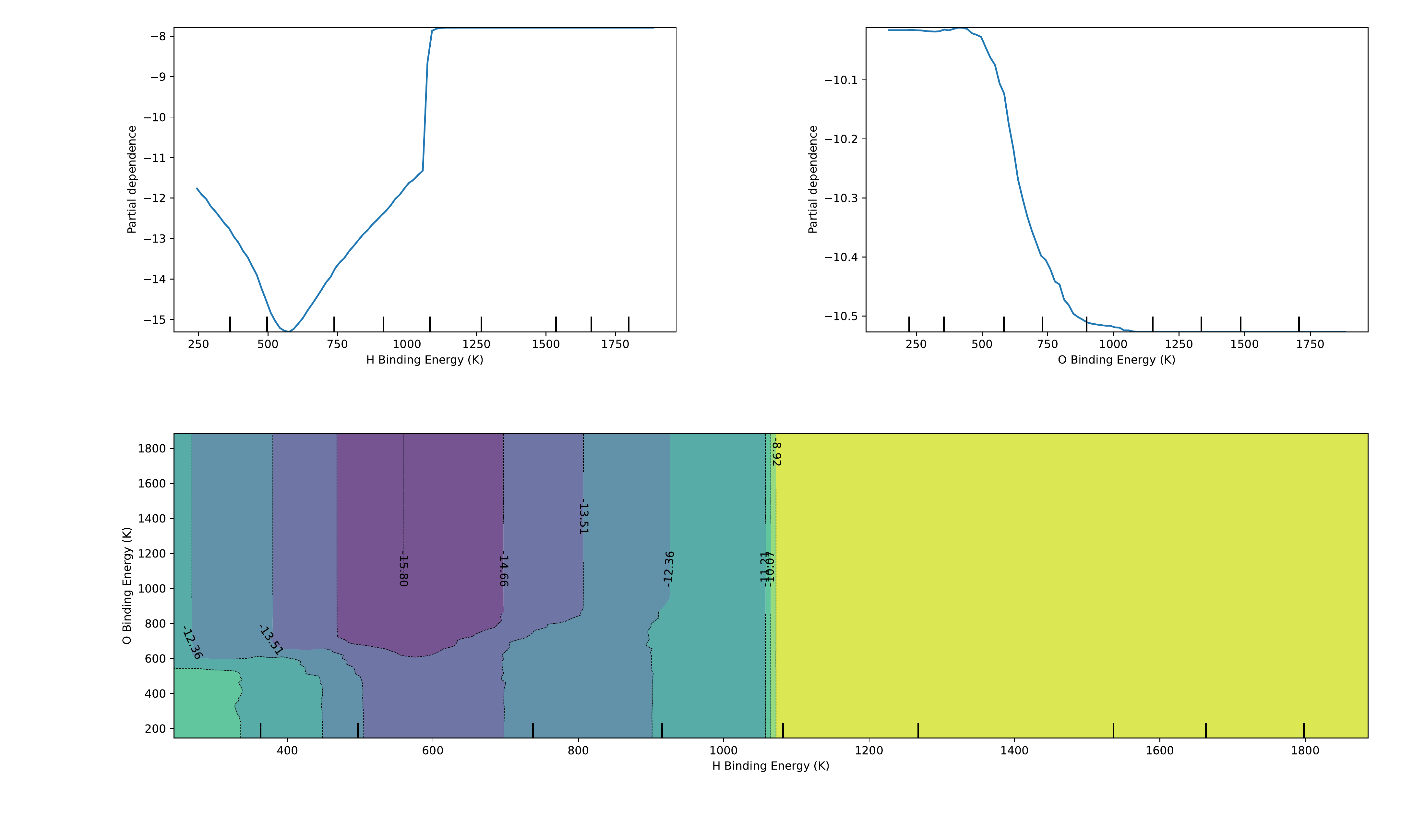}
\caption{Top: A plot of the 1-D partial dependence plots of the binding energies of H and O for CO. The partial dependence represents the expected value of the log-abundance of CO as a function of the variable in features, marginalised over all other features. Bottom: A 2-D partial dependence plot for the binding energies of H and O. Yellow represents the region with the highest abundance of CO.}
\label{CO partial dependence plots}
\end{figure*}

We can also consider the exact nature of the relationship between the features and the final abundance. To do this, we consider the partial dependence of specific variables relative to the output variable. The partial dependence is defined as the marginal effect of one or several features on the output of a machine learning model\cite{Friedmann_PDP,molnar2022}. To demonstrate the utility of the partial dependence, we consider \ce{H2O} and CO. Both of these molecules are largely dependent on two binding energies: that of H and O. We plot their 1-D and 2-D partial dependences in Figures \ref{H2O partial dependence plots} and \ref{CO partial dependence plots}. Note that the y-axis of the 1-D plots are simply the log-abundance of the respective species. 

We observe that for water, there is a small area of parameter space in which the abundance peaks. This roughly matches the maximum-posterior hydrogen binding energy value obtained in Figure \ref{bayesian_inference_plots}. Despite the oxygen's binding energy being the second most important feature, we observe that over the range of binding energies considered,  it has far less impact in changing the obtained water abundance. Even so, the parameter favoring binding energies lower than $\sim$1000K for oxygen is consistent with the posterior for the inverse problem.

We can make a similar comment about carbon monoxide. The abundance peaks for hydrogen binding energy values greater than 1100K. This makes sense, as having too low a binding energy for hydrogen would result in CO being hydrogenated efficiently. For binding energies above 600K for oxygen, we notice a slight decrease in the abundance of carbon monoxide.

\section{Conclusions}

In this article we focus our attention on the estimation of binding energies, key parameters in the interaction among surface reactions in ice. We use three statistical approaches to estimate binding energies, prioritise future ice species to be observed, and to understand better the non-linear relationship between binding energies and abundances of such species. Our conclusions can be summarized as followed:
\begin{itemize}
    \item As in our previous work, we find that Bayesian inference can be a very useful tool to constrain binding energies. However further ice observations are needed in order to reduce the variance of the distributions. 
    \item Indeed, the MOPED algorithm can  help towards the prioritization of such observations. As in previous work, we find that solid H$_2$O, CH$_4$, NH$_3$, H$_2$S, SiH$_4$, CO and H$_2$CO are the most important species to observe; surprisingly ice observations of CO$_2$, CH$_3$OH and HCN are not amongst the most significant species. 
    \item Using SHAP we establish the key relationships between binding energies and the abundances of the ice species. For example, we find that for water  and CO the key parameter is the hydrogen binding energy, and to a much lesser extent the oxygen one. Prioritizing which binding energies are keys for the potentially observable species may be of use in prioritizing experiments and calculations of such energies to reduce their errors.  
    
\end{itemize}

Probabilistic methodologies as well as Machine Learning  methods have  now started to be used to solve  astrochemical problems. As larger chemical reaction networks and more complex models are being employed in astrochemistry,  statistical methods and machine learning (ML) techniques will become ever more necessary in order to reduce the uncertainty in such networks. 

\section*{Author Contributions}
Johannes Heyl: Conceptualization, Data curation, Formal analysis,
Validation, Writing – original draft, Writing – review \& editing. 
Serena Viti: Conceptualization, Data curation, Formal analysis, Writing – original draft, Writing – review \& editing. Gijs Vermariën: Writing – review \& editing

\section*{Conflicts of interest}
There are no conflicts to declare. 

\section*{Acknowledgements}
J. Heyl is funded by an STFC studentship in Data-Intensive Science (grant number ST/P006736/1). This work was also supported by European Research Council (ERC) Advanced Grant MOPPEX 833460. 


\balance


\bibliography{rsc} 
\bibliographystyle{rsc} 

\end{document}